\documentclass[preprint2]{emulateapj-rtx4}
\usepackage{graphicx,amsmath,url}
\usepackage{bm}
\usepackage{multirow}
\graphicspath{{./fig/}{./png/}}

\newcommand{\EQ}{\begin{equation}}
\newcommand{\EN}{\end{equation}}
\newcommand{\EQA}{\begin{eqnarray}}
\newcommand{\ENA}{\end{eqnarray}}

\newcommand{\Eq}[1]{Equation~(\ref{#1})}
\newcommand{\Eqs}[2]{Equations~(\ref{#1}) and~(\ref{#2})}

\newcommand{\App}[1]{Appendix~\ref{#1}}
\newcommand{\Sec}[1]{Section~\ref{#1}}

\newcommand{\Fig}[1]{Figure~\ref{#1}}

\newcommand{\Tab}[1]{Table~\ref{#1}}

{}
{}
{}

{}
{}
{}
{}
{}
{}
{}
{}
{}
{}
{}
{}
{}
{}
{}
{}
{}
{}
{}
{}

{}
{}
{}

{}

{}
{}

\newcommand{\vv}{\mbox{\boldmath $v$} {}}

\newcommand{\kk}{\bm{k}}

\newcommand{\xx}{\bm{x}}

\newcommand{\uu}{\mbox{\boldmath $u$} {}}

\newcommand{\ii}{{\rm i}}

\def\St{\mbox{\rm St}}

\newcommand{\parr}[2]{\frac{\partial #1}{\partial #2}}

\newcommand{\aaa}{\mbox{\boldmath $a$}{}}{}

\newcommand{\mMM}{m_{\text{max}}}
\newcommand{\mmm}{m_{\text{min}}}
\newcommand{\mmp}{m_p}
\newcommand{\VV}{\bm{V}}
\newcommand{\Steff}{{St'}}

\begin{document}

\title{Turbulence induced collisional velocities and density enhancements: large inertial range results from shell models}
\author{Alexander Hubbard$^{1,2}$}
\affiliation{
$^1$ Max Planck Institut f\"ur Astronomie, K\"onigstuhl 17, D-69117
Heidelberg, Germany \\
$^2$ Department of Astrophysics, American Museum of Natural History, 79th St.\ at Central Park West, New York, NY 10024-5192, USA\\
}
\email{alex.i.hubbard@gmail.com}

\begin{abstract}
To understand the earliest stages of planet formation, it is crucial to be able to predict
the rate and  the outcome of dust grains collisions, be it sticking and growth, bouncing, or fragmentation.
The outcome of such collisions depends on
the collision speed, so we need a solid understanding of
the rate and velocity distribution of turbulence-induced dust grain collisions.
The rate of the collisions depends both on the speed of the collisions and the degree of clustering
experienced by the dust grains, which is a known outcome of turbulence.
We evolve the motion of dust grains in simulated turbulence,
an approach that allows a large turbulent inertial range making it possible to investigate the effect of turbulence
on meso-scale grains (millimeter and centimeter).
We
find three populations of dust grains: one highly clustered, cold and collisionless; 
one warm; and the third ``hot''.  Our results can be fit by a simple formula, and predict both
significantly slower typical collisional velocities for a given turbulent strength than previously considered,
and modest effective clustering of the collisional populations,
easing difficulties associated with bouncing and fragmentation barriers to dust grain growth.
Nonetheless, the rate of high velocity collisions falls off merely exponentially with relative velocity
so some mid- or high-velocity collisions will still occur, promising some fragmentation.
\end{abstract}

\keywords{turbulence  -- planetary systems: protoplanetary disks -- planets and satellites: formation}

\section{Introduction}

Collisions between dust grains in protoplanetary disks is a key process in planetesimal formation,
and so has seen a significant amount of study, both theoretical \citep{V80,M91,CH03,Y05,OC07} and
numerical \citep{DD05,J07}.
More, the results of collisions between artificial dust grains as a function of relative velocity
can be directly observed 
in laboratory experiments \citep{BW08, G10}.
These collisions can lead to sticking and growth of dust grains, or to fragmentation, 
repopulating the smallest sizes of dust grains.  On the other hand, laboratory experiments
suggest that the null-result of a collision, bouncing, poses a significant threat to dust growth beyond the cm scale
\citep{Z10}.

The gas component of protoplanetary disks is believed to be turbulent,
and the effect of this turbulence on the collisions between dust particles
entrained in the flow can be reasonably estimated analytically,
for example as done in the line of work starting with \citet{V80} and more recently elaborated on by
\citet{OC07}.  Such analyses give the unsurprising result that dust collisional velocities
are comparable to the turbulent velocities of the gas on scales set by the properties of the dust grains.
These estimated velocities are, however, expected to be large.  The turbulence in protoplanetary disks
is often invoked to explain the disk's accretion on the mega-year time-scales observed
\citep{SS73,DiskLifetime}, and the turbulent
motions required to achieve such a feat are expected to drive dust collisions that
destroy the participants or merely result in bouncing \citep{W01,G10, Wettlaufer10}.  The existence of
the bouncing and fragmentation velocity cut-offs means that
it is important to determine not merely the rate of collisions, but also the collisional probability
distribution as a function of the relative velocity, because outlying events could cross the bouncing
barrier \citep{Z10}.

Another important behavior of particles in turbulence is ``preferential concentration''
\citep{Maxey87,F94,C01,T09},
where inertial particles are ejected from regions of high vorticity due to centrifugal
forces, and accumulate in regions of high strain.  This has been hypothesized to increase
dust collision rates by creating local dust density fluctuations \citep{P11}, as well as possibly
leading to dust drag on the turbulence.  The latter can cause a streaming-instability
that further enhances the local dust density \citep{Y05,J07}.  Beyond the effect on
dust grain collisions rates, such density enhancements can provide a rapid route
to gravitational instabilities and the resulting collapse into planetesimals
 \citep{J06}.
This ejection of inertial particles from turbulent (intermittent) vorticies, which will be discussed
in this paper, should be distinguished from the dust-trapping property of
anti-cyclonic long-lived vortices, which are large enough to feel Coriolis forces \citep{BS95,J04}.

Moreover, the inertial range of the turbulence in these systems is expected to be
quite broad, easily five to six orders of magnitude in k-space or four orders of magnitude
in turbulent turn-over time, as their fluid Reynolds numbers are large.  This means that there
will be dust grains well embedded in the turbulent cascade, too small to notice that
they are trapped within huge eddies, yet too large to be meaningfully affected by the smallest
scale turbulence.
These well embedded particles are precisely the most important ones, since they include the
millimeter--centimeter sized dust grains that populate the fragmentation and bouncing
barriers. A resolution
on the order of several thousand cubed would be required to fully simulate a system 
evolving the full Navier-Stokes equations with an upper double digit inertial range, beyond currently available resources.
Further, as we will show, turbulent clustering implies that very small particle separations must be considered
to extract useful data.  The resolution required for performing a full hydrodynamical simulation is, again, on the order
of several thousand cubed (\Sec{Results:Clustering}).

\cite{Carballido10} explicitly run into the above-mentioned problems
of limited inertial range and low spatial resolution.
We resolve these problems by putting particles into a set of 
turbulent cascade models
known as \emph{shell models} (\citealt{shellbook}, chapter 3), which
allows us determine the clustering and collisional velocity probability
distributions that can be expected for large inertial ranges (up to $256$) while resolving
small particle separations.
This approach of using synthetic turbulence was also used by \cite{B05}, although
the details of the projection into real space differs.
Our work also differs by our focus on the collisional velocities as a key diagnostic, and we find
noticeably lower collisional velocities than those estimated analytically by \cite{V80} and 
following.  Further, unlike those works, our results are not well approximated by a single
effective collisional velocity, but require consideration of a velocity probability distribution.
While our clustering results are similar to those of \cite{P11}, simultaneously considering
the collisional velocities allows us to distinguish between multiple particle collisional populations,
which changes the conclusions about the effects of the strong clustering that is observed.

In \Sec{num} we
define our numerical methods and in \Sec{IA} we use them to investigate a particular base case to
find the physical behavior.  In \Sec{InertialRange}
we investigate systems with different inertial ranges and synthetic turbulence implementations.
We compare our results with previous work in \Sec{compprev} and conclude in \Sec{conclusions}.
We list our main variables, parameters, and diagnostics in Table \ref{var}
and discuss the definition and use of the Stokes number in \App{compstokes}.

\section{Numerical setup}
\label{num}

We use the {\sc Pencil Code}\footnote{http://pencil-code.googlecode.com}
to track the motion of particles in synthetic turbulent cascades.
Our synthetic turbulence derives from shell models: we consider spatial Fourier transforms of the 
turbulent velocity field (from $\xx$ to $\kk$),
keeping the modes that fit in a limited range of $|\kk|$ which defines our considered subsection of the
inertial range.  We then coarse-grain the sphere of $\kk$ vectors
into logarithmically spaced shells in $k$-space.  As such, we consider only the energy spectrum of the turbulence.
We will use four different models.  Three follow an exact Kolmogorov power spectrum
 (\citealt{K41}, see \Sec{pureKol}) and differ
only in the implementation of their time variability, if any.  The fourth model, discussed in \Sec{GOYmodel},
combines the spatial approach of \Sec{pureKol} with an energy spectrum derived from a GOY model,
which solves a set
of equations for the turbulent energy cascade.
However, it should be noted that the magnitude of the velocity
field in a GOY model is strongly fluctuating in time, making the choice of dimensionless quantities
less obvious.
A first study with a time-independent energy spectrum
is needed to establish the form of turbulent collisional velocities and is closer to existing analytical work.

Since we do not evolve the Navier-Stokes equations to determine the gas motion, we do not need a grid or a sub-grid model:
this allows for greater resolution.
We evolve
$4 \times 10^6$ particles in a periodic box of size $(2 \pi)^3$.  
Our setup
does not allow for the study of dust backreaction on turbulence (preventing, for example, the streaming
instability), but does allow us to examine dust pairs at very small relative separation.

\subsection{Flow field}
\label{pureKol}

Each coarse-grained shell is associated with a fluid velocity and turnover time and indexed
with $m$ which runs from $\mmm$ to $\mMM$ with
\begin{align}
&k_m=2^m, \\
&\tau_m=1/v_m k_m.
\end{align}
In a Kolmogorov spectrum, the velocity obeys the power-wave
\EQ
v_m = v_0 k_m^{-1/3}; \label{vkol} \\
\EN
this generates the dashed/red velocity spectrum in \Fig{goy_fig} by construction.
Our choice of $\mmm$ and $\mMM$ is limited by available numerical resources
and by the constraints of the periodicity of the numerical box.
We will vary them for different simulations as the choice determines
the effective inertial range of our turbulence-simulate.  This allows us to find the effective
inertial range large enough that the particle response becomes scale free (\App{compstokes}).

While \Eq{vkol} defines the amplitude of the velocity on spatial scales $k^{-1}$, we need an actual velocity field to evolve
our particle positions.  To obtain it, we associate each shell $m$ with three $(\kk_{mn},\hat{\vv}_{mn})$ vector
pairs indexed by $n$.  These vector pairs characterize contributions to the gas velocity directed along $\hat{\vv}_{mn}$, 
and varying along $\kk_{mn}$ (for shell $m$, $|\kk|=k_m$).
We impose $\kk_{mn} \perp \hat{\vv}_{mn}$, so that the flow is incompressible.
Further, the $\kk_{mi}$ and $\hat{\vv}_{mi}$ vectors for pair $i$ of shell $m$
are approximately perpendicular to their counterparts in the other two pairs of that shell.
This formulation (3 quasi-perpendicular
vectors for both $\kk$ and $\hat{\vv}$ for each shell in $k$-space) allows us to
span both $k$-space and velocity space in each shell separately while
simultaneously avoiding strongly preferred directions.  The projection of the velocity  $v_m$ associated with shell $m$
onto the three vector pairs is done through the introduction of a unit vector $\aaa_m$.  The three components of $\aaa_m$
define the relative excitation of the three possible vector pairs.

The gas velocity is then given by
\begin{align}
\VV(\xx,t)&=\Sigma_{m=\mmm}^{\mMM} \Sigma_{n=1}^3 \nonumber \\
&2^{1/2} a_{mn}(t)  v_m \hat{\vv}_{mn} \cos [\kk_{mn}\cdot \xx+\phi_{mn}(t)], \label{vgmodel}
\end{align}
where the outer sum is over shells in $k$-space, the inner sum is over the trio of ($\kk$,$\hat{\vv}$) vector pairs,
and $\phi_{mn}$ is a phase shift included to avoid overlap at $\xx=0$.  Note that $a_{mn}$ is a scalar (the n-th component of
$\aaa_m$, itself a unit vector)
while $\hat{\vv}_{mn}$ is the $n$th unit vector associated with the $m$th shell.
 Time dependence of the flow is allowed
in three ways.  First, the projection onto the $\kk$ and $\hat{\vv}$ vectors can be quenched
($\aaa_m$ and $\phi_{mn}$ held constant); these runs are named with a ``Q''.  Second, we
can vary the unit vectors $\aaa_m$ for each shell
through a random walk on timescales
$\tau_m$, with $\phi_{mn}$ held constant.  This is our default method, with runs denoted by a ``B''.
Third, we allow the various phases $\phi_{mn}$ to vary as random walks while
the $\aaa_m$ vectors are held constant in time; these runs are named with a ``P''.
Note that the GOY model inherently includes phase variation by evolving a complex value for $v_m$ (see
\Sec{GOYmodel}).

We will use the subscripts $ls$ (large scale) for $\mmm$ and $ss$ (small scale) for $\mMM$.
This approach of using an artificial velocity field is similar to that of \cite{B05}.
Our prescription is closer to that used in previous analytical studies.  It does however limit us to only considering
turbulence in the inertial range as we have no means of tracking behavior and the energy injection or viscous
dissipation scales.

Our box size is $L=2\pi$ and our boundary conditions are periodic,
so the minimal $k$ is $k=1$ ($m=0$).  The largest $k$ we will consider is $k=256$ ($m=8$).
In practical terms, assuming an energy-carrying scale at $k=4$ and a dissipative scale at 
a quarter of the Nyquist wave number one would require approximately $8000^3$ grid points to
directly simulate the turbulence for our largest inertial range (as compared to the very high
resolution of $2048^3$ in \citealt{B10dispersion}).
The benefit here can be seen in Figure 1 of \cite{Carballido10}, where the inertial range is, perhaps, $10$.
Moreover, in their Figure 3, the difference between the solid lines (assumed large Kolmogorov inertial range)
and dotted lines (actual evolved turbulence) shows how large the effect on the particle motions of the limited
inertial range actually is.

\subsubsection{GOY model}
\label{GOYmodel}

The Gledzer, Ohkitani and Yamada (GOY) model \citep{GOY} solves an evolution equation for the energy in differing shells under
the assumption that only shells adjacent in $k$-space interact (to generate a cascade rather
than non-local effects): shells $n$, $n+1$ and $n+2$ can interact.  We note that we solve the following
system for a larger range of shells than are included in the synthetic turbulence that is applied to the particles: the choice
$m=0$ ($k=1$) is associated with $n=4$ in what follows.  We follow $N=22$ shells, using a slaved Adams-Bashforth
scheme \citep{P93,M04}.
Under this assumption we can write
\begin{eqnarray}
\left(\frac{d}{dt} + \nu k_n^2\right)&&v_n=\ii\left( a_n v_{n+1} v_{n+2}+\right.  \label{GOYeq} \\
               &&\left.  b_n v_{n-1}v_{n+1}+c_m v_{n-1}v_{n-2} \right)^*+f_n, \nonumber
\end{eqnarray}
where $v_n$ is the complex velocity associated with shell $n$, $\nu$ is a viscosity
and $f_n=0$ for all $n \neq 1$ is the forcing term.
The choice of constraints of
energy and helicity conservation lead to
\EQ
a_n=k_n,\ b_n=-k_{n-1}/2,\ c_n=-k_{n-2}/2,
\EN
with boundary conditions on the largest and smallest pairs of shells ($n=0,1,N-1,N$) of
\EQ
b_1=b_N=c_1=c_2=a_{N-1}=a_N=0. 
\EN
The GOY shell model has a 3-cycle with shell index \citep{K95},
which we filter using
\EQ
|v_n^{\text{eff}}|=|\text{Im}(v_{n+2}v_{n+1}v_n-v_{n-1}v_nv_{n+1}/4)|^{1/3}.
\EN
This model inherently provides time variation for the phase through the complex nature of $u_n$.
We combine this with \Eq{vgmodel} by replacing $v_m$ with $|v_m^{\text{eff}}|$, setting $\phi_{mn}=\arg(v_m)$
and allowing $\aaa_m$ to vary as previously described.

\begin{figure}[t!]\begin{center}
\includegraphics[width=\columnwidth]{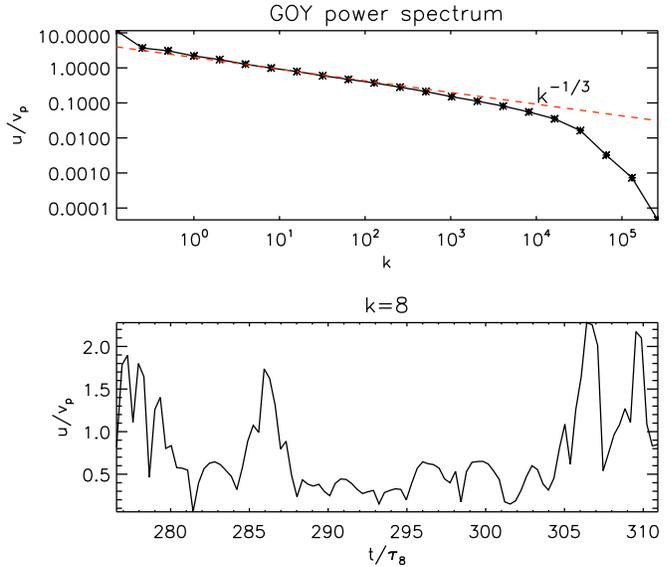}
\end{center}\caption{Velocity spectrum and time series for the GOY model.  In the upper panel,
we see that the power law of the GOY model is steeper than a Kolmogorov $-1/3$ law, closer to
$-0.4$.  In the lower panel, we show the velocity time series for the $k=8$ shell, with time
normalized to that shell's own turn-over time.
\label{goy_fig} }
\end{figure}

In \Fig{goy_fig} we show the velocity spectrum and a sample shell velocity time series.  We define a
characteristic shell velocity $\overline{u}_m$ through the time average of $|u_m|$ and $\tau_m=
1/\overline{u}_m k_m$, but this is not as perfectly defined as in the fully imposed Kolmogorov case.
The GOY model velocity spectrum is also slightly steeper than a Kolmogorov one,
a consequence of its intermittency.

\subsection{Particles}
\label{Sec:Particles}

\begin{figure}[t!]\begin{center}
\includegraphics[width=\columnwidth]{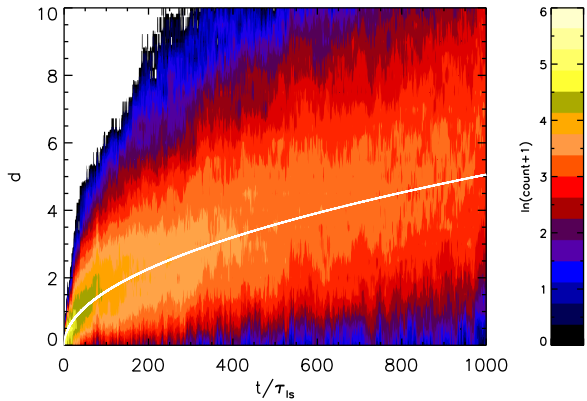}
\end{center}\caption{
Log scale histograms of the total distance $d$ traveled (in box units, box length $2\pi$) for 1000 particles as a function of time for a run with
 $\mmm=2$, $\mMM=5$.  Overplotted is
$c k_{ls}^{-1} (t/\tau_{ls})^{1/2}$ where $c=2/3$.
\label{histogram} }
\end{figure}

The motion of 
inertial particles in a fluid is determined by the friction between the particles and the fluid.
The resulting drag force entrains the particles along the fluid motion.
Particles with a finite friction time $\tau_p$ (subscript $p$ referring to particles) are referred to as
``inertial'' and their motion deviates from that
of the gas as long as the particles are not neutrally buoyant (ie. as long as $\rho_{\text{particle}} \neq \rho_{\text{gas}}$).
In this work we assume that $\rho_{\text{particle}}\gg \rho_{\text{gas}}$ which extremely well satisfied
for protoplanetary disks.  This allows us
to neglect pressure forces on the particles.
We initialize our particles with a frictional stopping time $\tau_p$, so that the equation for
a particle's velocity $\uu_p$ is determined by
\EQ
\parr{\uu_p(t)}{t}=-\frac{\uu_p(t)-\VV(\xx_p,t)}{\tau_p}, \label{dupdt}
\EN
where $\VV(\xx_p,t)$ is the gas velocity at the particle's position $\xx_p$.
It is this deviation of the particle velocity from that of the gas
that allows particles to collide even in incompressible flows.

When we evolve our particles in the velocity field given by \Eq{vgmodel}, we find a
particle velocity dispersion
\EQ
\langle |u_p^2|\rangle^{1/2} \simeq \frac{v_{\mmm}}{\left(1+\tau_p/t_{ls}\right)^{1/2}},
\EN
that matches the results expected since \cite{V80}.  Further, in \Fig{histogram} we show log scaled histograms
of the total distance $d$ traveled in box length units (box side=$2\pi$, bin size=$0.5$)
for a population of $1000$ particles in our base setup (see \Sec{IA}).  The time axis is scaled to the turnover
time of the largest scale turbulence (time between samples $0.2\tau_{ls}$).  We overplot
$0.67 k^{-1}_{ls} (t/\tau_{ls})^{1/2}$, which has the form and approximate scale of a random
walk on the length and time scales of the largest scale turbulence.  We can see that the particles'
displacements are well fit by the random walk expected to be generated by the
largest scale turbulent motion.

In this work we consider only particles with identical stopping time $\tau_p$ set to the turnover time of the shell $\mmp$
with $\mmm < \mmp<\mMM$.
We will therefore refer to $k_{\mmp}$ and $v_{\mmp}$ as $k_p$ and $v_p$.  In the limit of an infinite inertial range,
this scale is the only scale, and so most of our values will be non-dimensionalized with respect to it.  However, 
we can investigate the dependence of the collisional velocities on both smaller and larger eddies by
turning on and off eddies of differing scales, i.e., by changing $\mmm$ and $\mMM$.

We are primarily interested in dust collisions in protoplanetary disks, which allows some simplifications.
The
large size of the Kolmogorov scale turbulence in protoplanetary disks  (at a minimum kilometer scale),
 along with the small size of the dust grains
(quite sub-meter for grains embedded well within the inertial range) means that
when considering the collisions of protoplanetary dust grains we must treat
our dust-grains as point particles, taking limits as the particle separation tends to zero.
This is a distinguishing
feature compared to atmospheric turbulence, whose Kolmogorov scale might be $1$mm \citep{S03,X08}, comparable
to rain drops or particulates.  Finally, the actual collision rate of dust grains in protoplanetary disks
is expected to be modest due to their dust grain number density: grains will experience multiple
friction times between collisions.  Accordingly, given our focus on protoplanetary disks
we do not treat collisional scattering of the grains.

It is traditional to non-dimensionalize the particles' stopping time to the Stokes number by
scaling it to the turbulent turnover time at the dissipation scale (or occasionally to the time
at the driving scale).
Unfortunately neither definition
fits our setup well, but one can define $St \sim (k_{ss}/k_p)^{2/3}$ as a conceptual equivalent to scaling
$\tau_p$ to the dissipation scale since $k_{ss}$ is the smallest scale included in our simulation.
Similarly,
$St \sim (k_{ls}/k_p)^{2/3}$
is a conceptual equivalent to scaling $\tau_p$ to the turbulent time at the driving scale.
However, in \App{compstokes} we explain why we do not consider either approach to be
conceptually important as
we explicitly hope to achieve large enough
inertial ranges that the results are independent of $St$ (either definition).
We do define our effective Stokes number according to the first definition above,
$\Steff \equiv (k_{ss}/k_p)^{2/3}$, for use when we need to scale diagnostics
to the smallest included scale.  While it is not a perfect apples-to-apples comparison,
this value should be used when comparing with
most other work (e.g~\citealt{B10dispersion,P11}; but not \citealt{OC07}, where the other
definition is called for).

\subsection{Run details}

\begin{table}[b!]
\caption{Runs \label{Table:runs}}
\centerline{\begin{tabular}{lcccc}
\hline \hline
 Run & $\mmm$  & $\mmp$ & $\mMM$ & Notes \\
 \hline \hline
\multicolumn{5}{c}{Basic Turbulence (\Sec{pureKol})} \\
 B-A &  $0$ & $4$ & $6$  & Smallest $\tau_p/\tau_{ls}$ \\
 B-B &  $0$ & $3$ & $5$ & \\
 B-LI2 &$ 0$ & $ 3$ & $ 7$ & Largest inertial range $k_7/k_0=128$ \\ 
 B-C &  $1$ & $3$ & $5$ &\\
 B-LI1 &$ 1$ & $ 3$ & $ 7$ & Large inertial range $k_7/k_0=64$ \\ 
 Base & $ 2$ & $ 3$ & $ 5$ & Subject of \Sec{IA} \\
 B-D &  $2$ & $3$ & $6$   &\\
 B-E &  $2$ & $3$ & $7$ & \\
 B-F &  $2$ & $3$ & $8$ & Largest $\Steff=10$ \\
 \hline
\multicolumn{5}{c}{Quenched Spatial Projection (\Sec{pureKol})} \\
  Q-A & $2$ & $3$ & $5$ &  \\
 Q-LI &$ 0$ & $ 3$ & $ 7$ & Largest inertial range $ k_7/k_0=128$ \\ 
\hline
\multicolumn{5}{c}{Varying Phase (\Sec{pureKol})} \\
  P-A & $2$ & $3$ & $5$  & \\
 P-LI &$0$ & $ 3$ & $ 7$ & Largest inertial range $ k_7/k_0=128$ \\ 
\hline
\multicolumn{5}{c}{GOY Turbulence (\Sec{GOYmodel})} \\
  G-A           & $1$ & $3$ & $6$  \\
  G-B           & $2$ & $3$ & $5$  \\
  G-C           & $2$ & $3$ & $6$  \\
  G-LI     &$ 0$ & $3$ & $ 7$ &  Largest inertial range $ k_7/k_0=128$ \\ 
\hline \hline
\end{tabular}}
\end{table}

In \Tab{Table:runs} we collect the details of the simulations we perform.  The total inertial range
is given by $2^{\mMM-\mmm}$, the effective Stokes number by $\Steff=2^{2/3(\mMM-\mmp)}$,
 and the range between the largest
scale turbulence and the particles is given by $2^{\mmp-\mmm}$.  In all cases the time-step used
to advance the particles was $0.1 \tau_{ss}$.
For runs using the GOY model, the GOY model is advanced for $3\times 10^7$ time steps before the introduction of particles,
to allow the turbulence to find its statistical steady state.  The GOY equations are run with their own time-step of
approximately $10^{-4}$ in code units (tweaked to fit an integer number of times within a particle's time-step); this shorter
time-step is for the numerical stability of the algorithm.
The particles are initialized with $u=0$ at random positions.

We generate the $\kk_{mn}$ and $\vv_{mn}$ vectors of \Eq{vgmodel}
by randomly selecting $\kk_{m1}$ from the set of vectors
with approximately appropriate length (the error decreases as the target $|\kk|$ increases) that fit in the periodic
box.  The vector $\vv_{m1}$ is generated as a random unit vector perpendicular to $\kk_{m1}$.  The vector $\kk_{m2}$
is then selected from the set of vectors of appropriate length that fit in the box and lie within $30$ degrees of $\vv_{m1}$.
Next we choose $\vv_{m2}=\kk_{m1} \times \kk_{m2}/|\kk_{m1} \times \kk_{m2}|$, and repeat the process for $\kk_{m3}$
and $\vv_{m3}$.  If no vector $\kk_{m2}$ or $\kk_{m3}$ can be found that satisfy the constraints, the process
is restarted.

\section{Analysis: Behavior}
\label{IA}

We begin by making a full analysis of the run ``Base'' (\Tab{Table:runs}, smallest inertial range) to extract the behavior of the system,
both with respect to clustering and collision speeds, and to find a fit formula that can be applied
in simulations of particle size evolution.  In \Sec{InertialRange} we will study how both increasing the inertial
range  and changing our turbulence model affects the behavior.

We will obtain our particle clustering and collisional velocity data by
analyzing during run time snap-shots taken every full turbulent turnover $\tau_{ls}$
for the largest eddy in the system; these snap-shots include every particle's position and velocity. 
The particle positions are mapped onto a coarse grid, and every particle pair within a critical separation (at most $0.2$
of the grid scale) is found.  We bin our particle pairs simultaneously linearly in separation $R$, using either $50$ or $100$ bins
depending on the run,
 and linearly in relative velocity $u$. We consider
full spheres, so for every $R$ we consider every particle pair with $|\xx_1-\xx_2|<R$.  A bin listed with velocity $u_{\rm{bin}}$
contains particle pairs with relative velocity $u_{\rm{bin}}-\triangle u<u<u_{\rm{bin}}$ where $\triangle u=u_{\rm{max}}/n_u$,
$u_{\rm{max}}$ is the largest considered particle pair relative velocity and $n_u=2000$ is the number of velocity bins.
For run Base, $\triangle u=4\times 10^{-4}$ in code units .
Particle pairs with relative velocities larger than $u_{\rm{bin}}$
are included in that bin, so it is discarded.
We will refer to the total number of pairs in bin $(R,u)$ of snapshot $t$ as
$N(R,u,t)$, with dropped $u$ implying summation over all
velocity bins and
dropped $t$ implying averaging over snapshots.

We took $529$ snap-shots from run Base, but as discussed below,
for most of the analysis we drop the first twenty, accordingly, $N(R,u)$ and
$N(R)$ are averaged over $509$ snap-shots.

\subsection{Clustering}
\label{IA-clustering}

\begin{figure}[t!]\begin{center}
\includegraphics[width=\columnwidth]{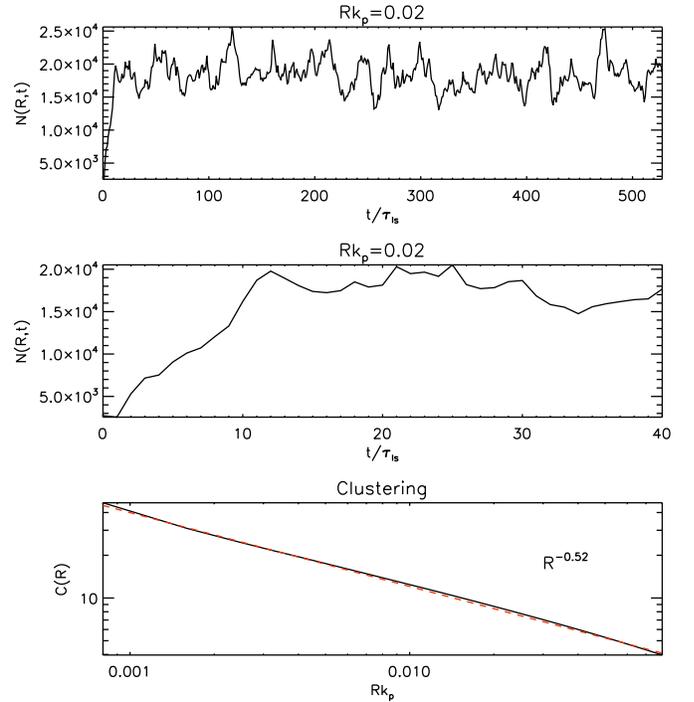}
\end{center}\caption{
Particle pairs $N(R,t)$ (unnormalized count) as a function of time for run Base.
Top panel: counts for a collisional sphere $R k_p=0.02$.
Middle panel: early time blow-up of top panel.
Bottom panel: Clustering C(R), with a power-law fit in red/dashed.
\label{cluster} }
\end{figure}

In the top two panels of \Fig{cluster} we show $N(R,t)$ 
as a function of time for a representative
maximal separation $R$.
We further define the clustering $C(R)$ as the ratio of $N(R)$ to the expected number of particle
pairs
\begin{align}
&\overline{N}(R) \equiv \frac 12 n_p \times \left(\frac{4\pi R^3}{3} \frac{n_p}{V}\right), \\
&C(R)=N(R)/\overline{N}(R), \label{Cdef}
\end{align}
assuming a spatially homogenous particle distribution, where $n_p$ is the number of particles
and $V$ is the volume of the box.  A value of $C(R)>1$ implies
that particles have been concentrated on length scales $R$, while $C(R)<1$ implies some
form of segregation or effective repelling.  The clustering $C(R)$ is the same as the
$g(St, r)$ of \cite{P11}, modulo the differences between $\Steff$ and their Stokes number.
Their length scale $\eta$ might be understood as our $k^{-1}_{ss}$, although
their turbulence at that scale is explicitly affected by dissipation, and so the energy is no longer following a
turbulent cascade.

The particles are initialized with zero velocity and random initial position, and in the top and middle panels 
of \Fig{cluster} we
see that the number of particle pairs takes more than ten largest-scale turn-overs to stabilize (more
than $16$ particle friction times), and still shows
significant fluctuations.  This long stabilization time is an important observation which helps explain 
results later in this paper:
there is significant structure in the particle distribution that takes time to develop, and studies
of particle collisions need to consider a significant time interval.  Unfortunately, including a history of
ten full largest-scale turnovers is beyond current or foreseeable analytical analyses.  Clearly particle positions
are quite correlated with one another.  In what analysis that follows, we discard the first twenty
snap-shots to allow the system to stabilize.

In the bottom panel of \Fig{cluster} we show $C(R)$ which
shows a clear power-law dependence on the maximal separation $R$.
We will denote by $\mu$ the power-law $C(R)\propto R^{-\mu}$, and by $R_1$ the intercept
where the power-law fit gives $1$.

\subsection{Average collision energy}

\begin{figure}[t!]\begin{center}
\includegraphics[width=\columnwidth]{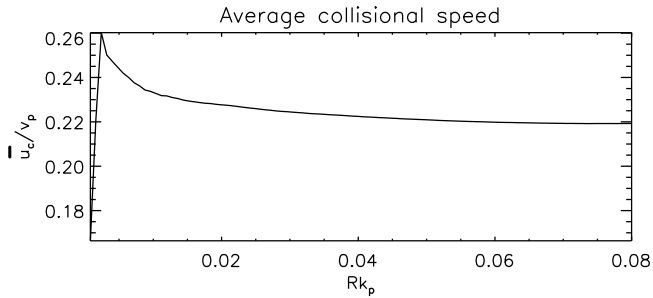}
\end{center}\caption{
Average collisional speed between particles as a function of maximal separation for run Base.
Note the small scale of the horizontal axis.
\label{avvc} }
\end{figure}

A first calculation of the rms-collisional velocity is shown in \Fig{avvc} where we plot
\EQ
\overline{u}_c(R) \equiv \left(\frac{\Sigma_u N(R,u) u^3}{\Sigma_u N(R,u) u}\right)^{1/2}.
\label{vc_formula}
\EN
The factor of $u$ in the denominator is needed to convert from particle pairs to particle
collision rates: because we are taking well-separated snapshots we need to weight high
velocity pairs more heavily.
A naive expectation would be that the collisional velocity decreases with decreasing
$R$, reaching a finite plateau at small separation, such as seen in \cite{B10v}.  This finite plateau
is made possible by the compressible nature of the particle ``fluid'', even if the turbulent gas itself
is incompressible.
In \Fig{avvc} we are already in this plateau, but
see an increasing collisional velocity with decreasing $R$ (except for the smallest $R$, where
our pair-counts are too low for reliability).

This perhaps surprising
result does not flow from poor statistics, but rather is a robust and understandable consequence
of the clustering seen in the bottom panel of \Fig{cluster}.  The clustering seen there
increases strongly with
decreasing $R$, implying a relatively numerous population of highly correlated particles.
As $R$ decreases, the relative particle-particle velocities in this highly correlated population
must decrease for the pairs to stay within the cluster radius: the smaller $R$, the 
lower the relative velocities of
the correlated population must be.
We might guess that the characteristic relative velocity
of this population is linear in $R$, giving a constant crossing-time.  Since the power-law fit
in \Fig{cluster} is weaker than $R^{-1}$, the contribution of this population
to the denominator of \Eq{vc_formula} decreases with decreasing $R$.
The velocity plateau at larger separations implies that the highly correlated population
has already ceased to contribute meaningfully to the numerator due to the extra factor of $u^2$.
Even at the largest separations $R$ we consider in \Fig{avvc}, the relative velocity of the highly correlated
population is low enough that this population contributes only negligibly to the total collisional power.
In effect, the highly correlated population is diluting the collisional energy when an
average is taken, and this dilution is decreased by considering smaller $R$.

Since we are considering
point particles, we must consider the limit $R \rightarrow 0$, and in that limit, the highly
correlated population
appears to play a dominant role in the pair counts (because $\mu > 0$), but contributes no collisions
(since $\mu<1$).  We accordingly define this collisionless population as ``cold'' (low relative velocities).
Unfortunately, numerical resource constraints prevent us from considering values of $R$
small enough that the cold population has ceased to contribute meaningfully to straightforward
calculations of the collision rate (the denominator of Equation \ref{vc_formula}).
Note that this is different from the expectations of, for example, \cite{P11}, where turbulent
clustering was hypothesized to lead to more common collisions by increase the number of nearby
potential collisional partners.  As we will see, however, there is a significant chance
that some level of clustering will persist, 
albeit not one that scales inversely with $R$.

\begin{figure}[t!]\begin{center}
\includegraphics[width=\columnwidth]{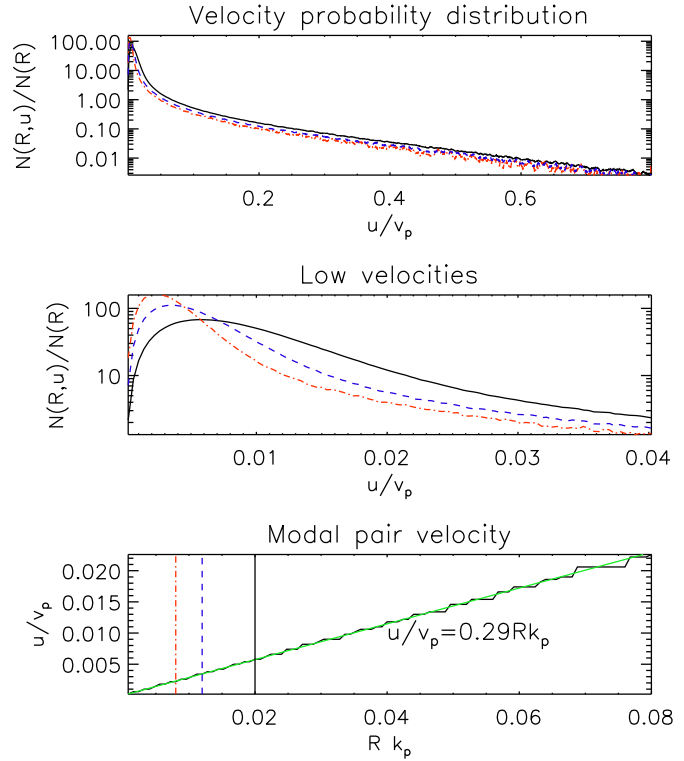}
\end{center}\caption{
Top and middle panels: particle relative velocity probability distributions for three different $R$
in run Base.  Black: $Rk_p=0.02$,
blue/dashed: $Rk_p=0.012$, red/dash-dotted: $Rk_p=0.008$; see also the top panels of \Fig{base_fits}.
Bottom panel: the velocity of the peak as a function of R along with a linear fit in green.  The vertical lines show the
position of the three
cases of the top two panels.
\label{paircounts} }
\end{figure}

The implications of this interpretation of \Fig{avvc} are seen in \Fig{paircounts}, where 
\EQ
N'(R,u) \equiv N(R,u)/\triangle u
\label{Nprime}
\EN
 is the density of pairs within separation
$R$ in velocity space (to convert between binned data and a smooth distribution).
The probability distribution is
seen to be strongly affected by a low velocity peak which moves to smaller velocity linearly with $R$,
to a limit of $0$ as $R \rightarrow 0$ (compare with \citealt{Carballido10}, Figure 5, which is lin-log instead of log-lin).
We define
\EQ
\omega \equiv u_{cold}/R, \label{omega}
\EN
as the measure of the dependence of the velocity of the cold population on separation,
where $u_{cold}$ is the modal pair velocity.

This collisionless behavior of the cold population does not mean that we expect no collisions.
As can be seen in the top panel of \Fig{paircounts}, $N(R,u)$ has a significant
high velocity tail, which can contribute ``real'' collisions.
A process to excise the cold, collisionless population and fit the remainder is discussed below.

\subsection{Fitting}

\begin{figure}[t!]\begin{center}
\includegraphics[width=\columnwidth]{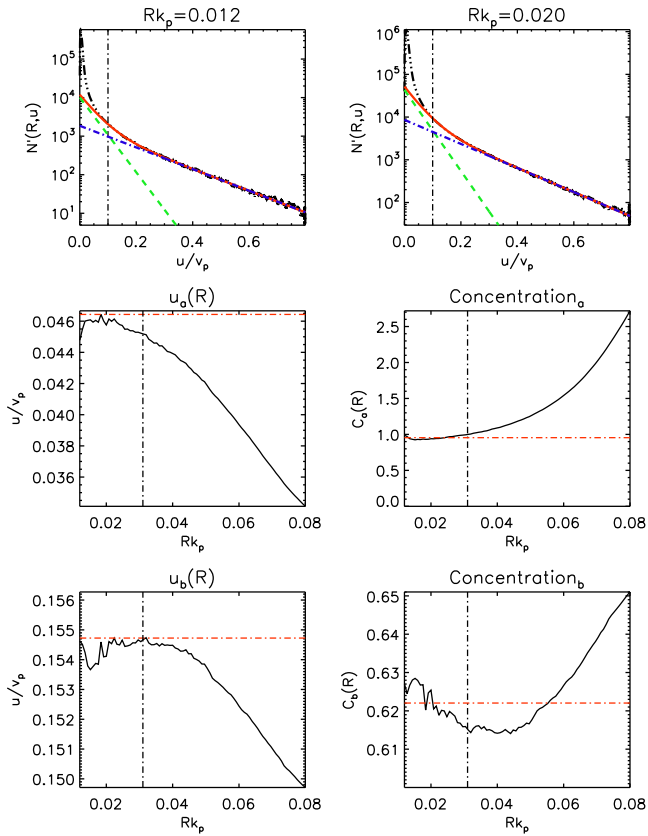}
\end{center}\caption{
Top panels: fits using \Eq{fitform} for run Base.  Black/dash-triple-dotted: data, red/solid: total fit, green/dashed: fit component $a$,
blue/dash-dotted: fit component $b$.  Only the region to the right of the vertical line
$u=u_c$ is fitted.
Left panels: $u_a$ and $u_b$, along with their maxima in the non-contaminated region to
the left of the vertical line $R=R_c$.
Right panes: $C_a$ and $C_b$, along with their means in the non-contaminated region to the
left of the vertical line $R=R_c$.
\label{base_fits} }
\end{figure}

We have failed to successfully fit the cold, clustered population with a tractable
distribution such as a Weibull distribution; however the tail of the pair-probability
distribution in velocity space is well fit by a simple exponential, and at more modest velocities
the addition of a second exponential improves matters.  Accordingly, we fit the particle-pair
probability distribution by
\begin{align}
&N'(R,u>u_c)/\overline{N}(R)=\frac{C_a}{u_a} e^{-u/u_a}+\frac{C_b}{u_b} e^{-u/u_b}, \label{fitform} \\ 
&u_c=0.1v_p \label{uc}
\end{align}
but only for $u>u_c$, where $u_c$ is a cut-off designed to exclude the cold population.
The fit formula \Eq{fitform} identifies two separate populations $a$ and $b$ with $u_a<u_b$.
While we do not have an analytical theory for the form of \Eq{fitform}, the quality of the high velocity
component $b$ can be seen from the overlap between the blue/dash-dotted curves (the higher
velocity component) and the black/dash-triple-dotted curve (data) in \Fig{base_fits}.
The fit for component $a$ to be less well constrained, but it gives a noticeable different
at lower relative velocities, seen in the difference between the blue/dash-dotted curve (only
component $b$), and the red/solid curve (both components, overlays the data outside of $u_c$).
We refer to component $a$, handling
the lower relative velocities, as the ``warm'' population, 
and the fit component $b$, handling the higher relative velocities, as the ``hot'' tail.

The parameters $C_a$ and $C_b$ then set the clustering of the effective densities of the warm 
and hot populations compared to the total population were it homogeneously distributed
(see Equation~\ref{Cdef}).
Finally, we define a contamination radius 
\EQ
R_c=0.2 u_{a0}/\omega \label{radcont}
\EN
 outside
of which the cold population is considered to contaminate the results, i.e., where
$u_{cold}$ from \Eq{omega} is comparable to $u_a$.
  The factor of $0.2$ is chosen ad-hoc,
but is inspired by the middle panels of \Fig{base_fits}; $u_{a0}$ is an approximate value for $u_a$.
The fit parameters in \Eq{fitform} are functions of $R$, but as seen in \Fig{base_fits} meaningful
limits $R\rightarrow 0$ can be taken.
The reported values of $C_a$, $C_b$, are averaged over the non-contaminated region, while
the reported values of $u_a$, and $u_b$ are the maximum value inside the non-contaminated region.

\subsection{Weights and errors}

Our reported fit coefficients $u_a, u_b, C_a$ and $C_b$ depend on the weighting scheme, the fitting region, 
and $u_{a0}$.  We weight with
\EQ
W=[1/N(R,u)]\times \left[1+\frac{2(u-0.7 u_{\rm{M}})(u-u_c)}{(0.7 u_{\rm{M}}-u_c)^2}\right],
\EN
where $u_{\rm{M}}$ is an upper velocity cut-off imposed when the number of pairs/bin approaches unity.
This weighting scheme weights the low velocity (warm) and high velocity (hot) regions more highly, allowing the two
populations to be distinguished by the fitting algorithm.
The values of $N'(R,u)$ in \Fig{base_fits} have a minimum $\sim 10$.  As we are averaging over $509$
snap-shots, with $\triangle u=4\times 10^{-4}$, that corresponds to $\sim 2$ pairs per bin which is
a consequence of our large number of velocity bins, $2000$.  To avoid
zeroes we have smoothed over $20$ bins in velocity space.
The effect of sampling noise can be gauged from the middle and bottom panels of \Fig{base_fits}
as the limit $R\rightarrow 0$ is taken with a corresponding reduction in pair count, especially
at high relative velocity.
As the fitting procedure is somewhat qualitative, we will be more interested
in the scale of the results, and any trends with varying $\mmm$ and $\mMM$, rather than in precision.
As we will discuss in \Sec{compprev}, our collisional velocities are factors of order five slower
than previous work, well outside any systematic or statistical errors.

\section{Analysis: Inertial Range, Turbulence Models}
\label{InertialRange}

\subsection{Clustering}
\label{Results:Clustering}

\begin{table}[b!]
\caption{Clustering diagnostics \label{Table:clust}}
\centerline{\begin{tabular}{lcccccccc}
\hline \hline
 Run & $\mmm$  & $\mmp$ & $\mMM$ & $\Steff$ & $\mu$ & $R_1 k_p$ & $\omega\tau_p$  & $R_c k_p$  \\
 \hline 
B-A       &  $0$ & $4$ & $6$ & $2.5$ & $0.65$ & $1.6$ & $0.29$ & $0.04$ \\
B-B       &  $0$ & $3$ & $5$ & $2.5$ & $0.57$ & $1.9$ & $0.27$ & $0.04$ \\
B-LI2 & $0$ & $3$ & $7$ &  $6.3$ & $0.39$ &$1.6$ & $0.55$ & $0.02$ \\
B-C       &  $1$ & $3$ & $5$ & $2.5$ & $0.55$ & $1.7$ & $0.23$ & $0.04$ \\
B-LI1 & $1$ & $3$ & $7$ &  $6.3$ & $0.38$ & $1.8$ & $0.55$ & $0.02$ \\
Base &  $2$ & $3$ & $5$ & $2.5$ & $0.49$ & $1.6$ & $0.29$ & $0.03$ \\
B-D       &  $2$ & $3$ & $6$ &    $4$ & $0.45$ & $1.5$ & $0.39$ & $0.02$ \\
B-E       &  $2$ & $3$ & $7$ & $6.3$ & $0.38$ & $1.2$ & $0.52$ & $0.02$ \\
B-F       &  $2$ & $3$ & $8$ &  $10$ & $0.38$ & $1.2$ & $0.76$ & $0.02$ \\
  Q-A & $2$ & $3$ & $5$ & $2.5$ & $0.34$ & $5.0$ & $0.51$ & $0.03$ \\
  Q-LI & $0$ & $3$ & $7$ & $6.3$ & $0.44$ & $13$ & $0.50$ & $0.03$ \\
  P-A & $2$ & $3$ & $5$ & $2.5$ & $0.81$ & $0.8$ & $0.21$ & $0.03$ \\
  P-LI & $0$ & $3$ & $7$ & $6.3$ & $0.40$ & $2.2$ & $0.39$ & $0.03$ \\
  G-A           & $1$ & $3$ & $6$ & $3.7$ & $0.36$ & $1.9$ & $0.14$ & $0.02$ \\
  G-B           & $2$ & $3$ & $5$ & $2.4$ & $0.40$ & $2.0$ & $0.16$ & $0.03$ \\
  G-C           & $2$ & $3$ & $6$ & $3.7$ & $0.33$ & $2.7$ & $0.14$ & $0.03$ \\
  G-LI           & $0$ & $3$ & $7$ & $5.9$ & $0.38$ & $1.8$ & $0.20$ & $0.02$ \\
\hline \hline
\end{tabular}}
\end{table}

In \Tab{Table:clust} we present the basic clustering data for our turbulence models,
across an array of inertial ranges and our turbulence models.
Parameters $\mmm,\mmp,\mMM$ and $\Steff$ define the included inertial range,
as well as an effective Stokes number.  The diagnostics $\mu$ and $R_1$ are the
power-law and intercept of the power-law fit to $C(R)$ (\Sec{IA-clustering}) while
$\omega$ defines the relative velocities of the cold population (Equation \ref{omega})
and $R_c$ defines the contamination radius outside of which the collisionless cold population
becomes indistinguishable from the collisional warm and hot populations.

There appears to be a modest decrease of the clustering parameter
$\mu$ with the number of shells with turnover times shorter than
the particle drag time $\tau_p$, as measured by $\Steff$.  However, it also appears to increase with the number of shells included
with turnover times longer than $\tau_p$.  Comparing runs B-E, B-F, B-LI1, and B-LI2, we feel
that we have reached a large enough inertial range to estimate that particles well embedded in turbulence
will experience clustering with $\mu \sim 0.4$.  This result matches the behavior with other turbulence models,
although run P-A is a clear outlier.
 
Another interesting column is $\omega \tau_p$.  This value scales
slightly slower than $\Steff$, and $\omega^{-1}$ is the crossing time
of a cold cluster.  Accordingly, the result is that the crossing time
of the cold cluster scales slightly slower than the turnover time
of the smallest included eddy, supporting the extension of the hypothesis
that the cold velocity is linear with separation (Eq. \ref{omega}) to an infinite cascade without
meaningful smallest scale.  This is further born out by the results for the other turbulence
models, particularly the GOY model, which see weak or no dependence of $\omega$ on $\Steff$.

A problematic column is $R_c k_p$.  This imposes the minimal spatial resolution required
to make a meaningful analysis of dust collisions by discarding the cold population.  In a
$2\pi^3$ box with $k_p=8$ (assuming an energy carrying scale of $k=4$),
a grid scale with $\triangle x=R_c$ requires a resolution
on the order of $1700^3$ even for our smallest inertial range.  This smallest range is within reach,
albeit barely \citep{B10dispersion,B10v}.

\subsection{Collisions}
\label{Analysis:Col}

\begin{table}[b!]
\caption{Collisional diagnostics \label{Table:col}}
\centerline{\begin{tabular}{lcccccccc}
\hline \hline
Run & $\mmm$  & $\mmp$ & $\mMM$ & $\Steff$ & $C_a$ & $C_b$ & $u_a/v_p$ & $u_b/v_p$ \\
 \hline
B-A       & $0$ & $4$ & $6$ & $2.5$ & $1.1$ & $0.46$ & $0.070$ & $0.21$ \\
B-B       & $0$ & $3$ & $5$ & $2.5$ & $1.2$ & $0.57$ & $0.056$ & $0.20$ \\
B-LI2 & $0$ & $3$ & $7$ &  $6.3$ & $1.7$ &$0.92$ & $0.069$ & $0.19$ \\
B-C       & $1$ & $3$ & $5$ & $2.5$ & $0.9$ & $0.42$ & $0.054$ & $0.18$ \\
B-LI1 & $1$ & $3$ & $7$ & $6.3$ & $1.5$ & $0.89$ & $0.069$ & $0.17$ \\
Base & $2$ & $3$ & $5$ & $2.5$ & $1.0$ & $0.62$ & $0.046$ & $0.15$ \\
B-D       & $2$ & $3$ & $6$ &    $4$ & $1.1$ & $0.83$ & $0.054$ & $0.17$ \\
B-E       & $2$ & $3$ & $7$ & $6.3$ & $1.1$ & $1.0$ & $0.057$ & $0.16$ \\
B-F       & $2$ & $3$ & $8$ & $ 10$ & $1.2$ & $1.1$ & $0.064$ & $0.17$ \\
 Q-A & $2$ & $3$ & $5$ & $2.5$ & $1.4$ & $1.4$ & $0.077$ & $0.48$ \\
 Q-LI & $0$ & $3$ & $7$ & $6.3$ & $2.8$ & $2.0$ & $0.13$ & $0.71$ \\
  P-A & $2$ & $3$ & $5$ & $2.5$ & $0.5$ & $0.25$ & $0.036$ & $0.098$ \\
  P-LI & $0$ & $3$ & $7$ & $6.3$ & $1.5$ & $0.37$ & $0.050$ & $0.134$ \\
 G-A           & $1$ & $3$ & $6$ & $3.7$ & $0.7$ & $0.31$ & $0.070$ & $0.22$ \\
 G-B           & $2$ & $3$ & $5$ & $2.4$ & $0.7$ & $0.43$ & $0.077$ & $0.32$ \\
 G-C           & $2$ & $3$ & $6$ & $3.7$ & $0.7$ & $0.31$ & $0.068$ & $0.20$ \\
 G-LI           & $0$ & $3$ & $7$ & $5.9$ & $0.8$ & $0.37$ & $0.076$ & $0.24$ \\
\hline \hline
\end{tabular}}
\end{table}

In Table \ref{Table:col} we collect fit data for our simulations, which should only be used for intermediate
and higher relative velocities ($u>u_c=0.1v_p$, see Equations \ref{fitform} and \ref{uc}).
Parameters $\mmm,\mmp,\mMM$ and $\Steff$ define the included inertial range,
as well as an effective Stokes number, while the diagnostics $C_a, C_b, u_a$ and $u_b$
are the coefficients in our fit formula \Eq{fitform}.  Coefficients $C_a$ and $C_b$ describe
the effective target number density seen by the warm and hot populations respectively, and
coefficients $u_a$ and $u_b$ represent the velocity scale of the two populations.

The $C_a$ and $C_b$ columns indicate
that the warm and hot populations, in the velocity regime where the fit applies, have individual effective densities comparable
to the density that the total dust population would have were it evenly distributed.  This does not strictly imply that
there is clustering of the collisional population, because many of the collisions are likely at low, unfitted velocities.
The results for $C_a$ and $v_a$ are the least useful, as they are most sensitive to the cold population, and $u_a<u_c$.
The results for $u_b$ are nicely only weakly sensitive to the set of included shells, while $C_b$ appears to
require a significant $\Steff \simeq 10$ to reach a limiting value.

Most of our diagnostics, such as $u_a, u_b, C_a$ and $R_1$ appear to be relatively insensitive to the
set of included shells, i.e., $(\mmm-\mMM)$ and $\Steff$.  It should nonetheless be noted
that since the diagnostic $u_b$ is crucial to the high velocity collisions which might result in fragmentation,
its weak sensitivity to $\mmm$ is still important.  Achieving converged values for
$C_b$, an important diagnostic, certainly requires $\Steff \gtrsim 6$, while the behavior of $C_a$ in the limit
of both larger and smaller eddies is not yet clear (bottom two lines of \Tab{Table:col}).  Accordingly, ranges
of $k_p/k_{ls}=8$ and $k_{ss}/k_p=16$ appear to be reasonable minimums.
Unfortunately, it appears that extreme resolution is required not merely to include an adequate inertial range,
but also to capture the details of particle clustering.

\begin{figure}[t!]\begin{center}
\includegraphics[width=\columnwidth]{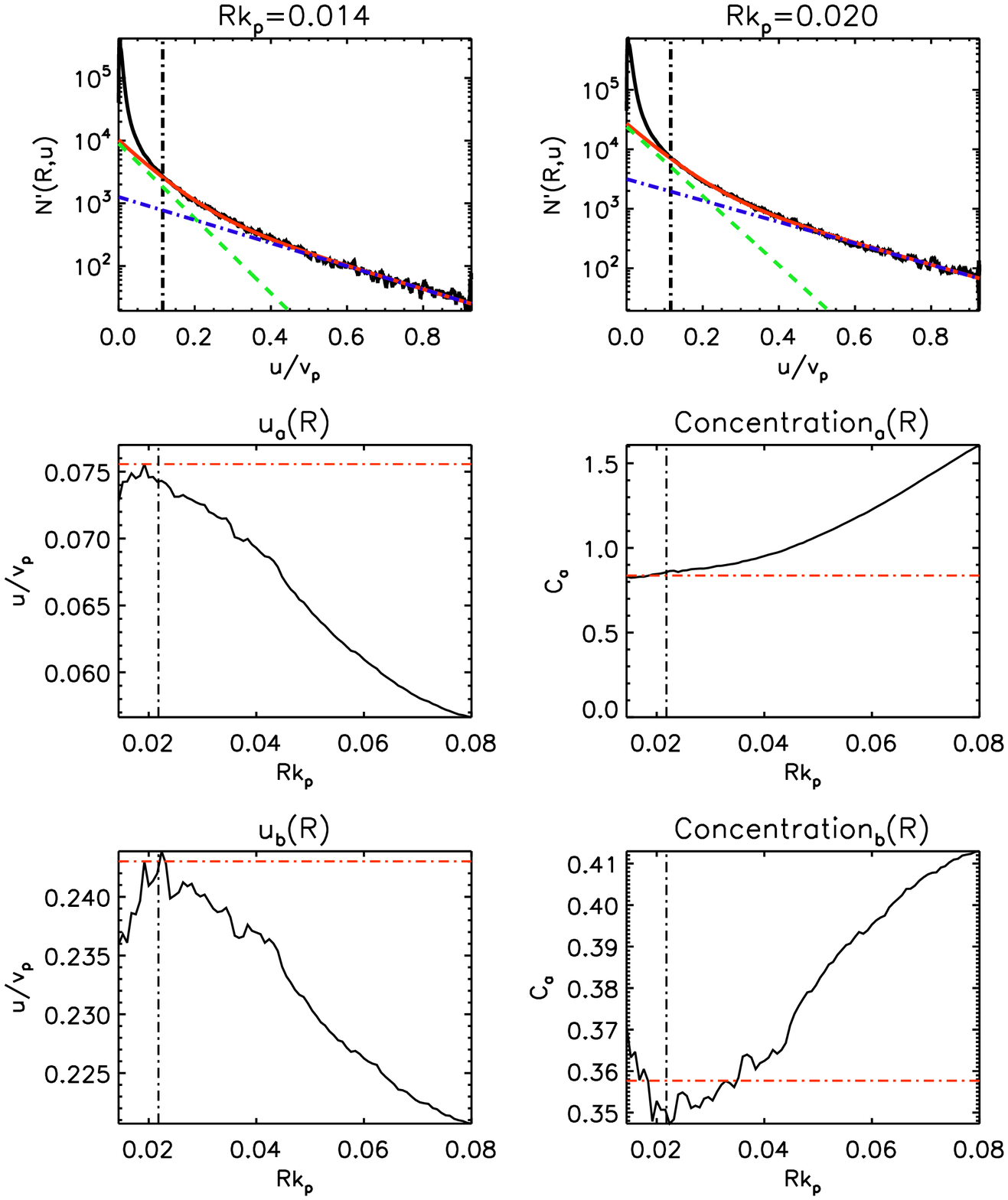}
\end{center}\caption{
As in \Fig{base_fits}, but here we show run G-LI.  Averaged over $112$ snap-shots, with
$\triangle u=4.64\times 10^{-4}$ in code units.
\label{goy_fits} }
\end{figure}

Comparing the results across different turbulence models,
a first observation is that they are well fit by \Eq{fitform}, as seen in \Fig{goy_fits}, albeit with differing parameters.  The
fully quenched model (runs Q-A and Q-LI) produces both significantly more concentrated
collisions (the $C$ coefficients), and significantly
higher collisional velocities.  This is not particularly surprising as there is nothing to destroy long time
correlations.  Putting the time variation of the spatial projection into the phase (runs P-A and P-LI) however results
in a significant drop in the collisional velocities.  Finally, the GOY model appears to fit
with very similar velocity parameters of our base model although the clustering is weaker.
This is somewhat surprising as the
turbulence is spiky, which might be expected to result in a stronger high velocity tail.  On
the other hand, it inherently includes phase variation.  A further unexpected detail of the
GOY model is that the smallest range of included shells ($\mmm=2$, $\mMM=5$) showed
the strongest high velocity tail ($u_b=0.32v_p$).  This may be due to the fact that the GOY scheme has an
intermittent energy cascade, and the generation of sub-vortices from larger vortices (a consequence
of localized energy spikes in $k$-space, through the coupling terms of Equation \ref{GOYeq}), if appropriately
followed by including an adequate inertial range to track the sub-vortices, will decrease particle correlations.

Finally, we suggest that \Eq{fitform} be used with parameters of the order of
 $C_a \sim 1.5$, $C_b \sim 1$,
$u_a \simeq 0.07 v_p$ and $u_b \simeq 0.2 v_p$, applying only when relative velocities
exceed the cut-off velocity $u_c=0.1 v_p$, i.e.,
\EQ
\frac{N(R,u>0.1v_p)}{\overline{N}(R)}=\frac{21}{v_p} e^{-14u/v_p}+\frac{5}{v_p} e^{-5u/v_p}. \label{fit!} \\ 
\EN
If one prefers the GOY model, the concentration parameters should be changed to $C_a \sim 0.8$ and $C_b \sim 0.4$.

\section{Comparison with previous work}
\label{compprev}

\subsection{Clustering}
\label{compclust}

Turbulent clustering has been a topic of interest in the protoplanetary community, and has been invoked,
for example, in the creation of chondritic meteorites \citep{C01}.  Our clustering diagnostics follow those of
\cite{P11}, rather than those of \cite{F94,Hogan01}.
The previous work
found evidence for strong clustering for particles with modest $St$, i.e., particles reasonably
well coupled to turbulence at the dissipation scale.  The clustering of \cite{P11}
was observed to decrease as their Stokes number became different from $1$.
Our results fit in magnitude with those of \cite{P11}, which scale up to $\mu\gtrsim 0.6$ (see their Figure 6).
However, our clustering is significantly stronger at high $St$ or $\Steff$.
Given the differences
in the fluid flows used (synthetic turbulence in our case, forced turbulence in \citealt{P11}), 
the similarities in behavior and scale of the exponent
are encouraging.  However, it isn't clear from the $\mu$ column in \Tab{Table:clust}.
that clustering will cease for large $\Steff$, as previously seen.
It should be
noted that previous simulations had limited inertial ranges, so their high $\St$ particles
are not fully embedded in the turbulence (see \App{compstokes}).

There is reason to expect
that clustering will be more effective for particles with $\St=1$ \citep{C01,Hogan01}.
However, we can imagine
evolving a system of two species of particles with $\St_1=1$ and $\St_2 \ll 1$.
Once a statistical steady-state is achieved, with the $\St_1$ particles more strongly clustered
than the $\St_2$ particles, we decrease
the fluid viscosity significantly so that in the new flow, $\St_1 \gg 1$ and $\St_2=1$ because the new dissipative scale is
smaller.  At that point, even if the $\St_2$ particles are more strongly clustered than the $\St_1$ particles,
it isn't clear why the $\St_1$ particles would be less clustered than originally.  Runs Base through B-F
in \Tab{Table:clust} match this thought experiment, and do show a drop of $\mu $ from $0.49$ to $0.38$
which appears to be a minimal value.  This weak decrease in $\mu$ may result from particle clusters being
disrupted by the additional, smaller-scale, eddies.

\cite{P11} hypothesized that clustering, especially if it exists for $St>1$,
might greatly increase the collision rates between particles by generating regions of enhanced
particle number density.  Our results allow us to contradict the straightforward version of that hypothesis: the clustering exists,
but the particle clusters are non-collisional.  However, once particle number densities get high enough that
the dust fluid density is comparable to the gas density, the dust drag has a significant back-reaction on
the gas.  An example is the streaming instability.  In protoplanetary disks,
gas orbits in a sub-Keplerian fashion because of the outwards pointing pressure force.  As a result,
particles, which would naturally orbit in a Keplerian fashion, feel a headwind.  Similarly to drafting
on highways or in bicycle races, clumps of particles can then form which are dense enough to back-react
on the gas \citep{C01,Y05, J07, Tornado}.

\subsection{Collisions}
\label{compcol}

Much of astrophysical turbulent dust collision work follows the approach of \cite{V80,M91,CH03};
see \cite{OC07} for a recent.  This approach defines two classes
of eddies, Class I large-scale eddies and Class II small-scale eddies.  The former 
have large velocities and time scales, and so can transport dust grains significant distances:
they dominate the turbulent diffusion of dust throughout a disk or atmosphere.
On the other hand, they change slowly enough in
both space (compared with dust stopping lengths) and time (compared with $\tau_p$)
that nearby dust grains, which could collide, see nearly identical
gas motion.  As a result, Class I eddies can affect the collisional behavior of dust grains only
slightly.  Class II eddies on the other hand vary rapidly compared to both the frictional
stopping time of the dust grains and their stopping length.  Accordingly, these eddies
can affect even nearby dust grains differently, driving collisions.  However, their short
time and length-scales mean that they provide only weak large-scale transport.

Dropping the
contribution of Class I eddies for collisions between identical dust grains,
\cite{OC07} find an rms collisional velocity of 
\EQ
u^2_{\text{Ormel}}=2 v_p^2. \label{OrmelV}
\EN
While this result is frequently quoted in terms of the Stokes number defined as $St=\tau_p \Omega_K$,
the above version is more general (see \App{compstokes}).
An important assumption is that the relative motion of particles approaches a finite limiting
value as their separation goes to zero, which is possible for inertial particles ($\tau_p \neq 0$)
whose motion must deviate from that of the gas.  Such behavior is seen in the direct numerical
simulations of \cite{B10v} and we believe
we understand the deviation from that assumption seen in our \Fig{avvc}.
This analytical approach cannot however handle very long time-correlations between
dust grains.

The result in \Eq{OrmelV} is different from our Equation \ref{fitform} in two ways.  First, it is a single number, which
does not suffice to describe our results, with their two velocity scales of the warm and hot
populations.  Our results also have merely exponentially falling probability tails with $u$, a much
broader distribution than, for example, a Maxwellian distribution.  They should therefore be treated as a
probability distribution.  Second, if one nevertheless uses \Eq{vc_formula} to extract a single rms-averaged
collisional velocity, we find
\EQ
u_{\text{our}}=\left[ \frac{6 \left(C_a u_a^2+C_b u_b^2\right)}{C_a+C_b}\right]^{1/2} \simeq u_{\text{Ormel}}/4,
\EN
using the suggested parameters of \Sec{Analysis:Col},
a significant decrease in the characteristic collisional velocities and a much larger decrease
in the turbulent collisional energies.  We attribute this difference in
predicted collisional velocities to
the high level of correlation we have identified, which not only creates the ``cold'' population,
but appears to slow all the collisions.

One can also compare 
\Fig{avvc} and the bottom panel of \Fig{paircounts} with
Figure 4 of \cite{Carballido10}, which is further limited by the sub-gridscale gas velocity interpolation.
Clearly, very high spatial resolution is required to extract actual collisional velocities from numerical
simulations, at least for
particles with identical stopping times.  
Our method provides large inertial ranges and high resolutions, whose necessity can be seen in their Figure 3, 
where the analytical result based on a full inertial range, the analytical
result based on the actual inertial range and the numerical result all disagree.
The resolution to begin to see the multiple populations we have identified in direct numerical simulations
of the full Navier-Stokes equation may be becoming available (for example the $2048^3$ simulation
of \citealt{B10dispersion,B10v} which is just adequate to include the inertial range assumed in run Base while resolving
$R_c$).
However the analysis has not been done in the same terms so it is not
clear whether they would see the effects we predict, or, if not, it would be because they contradict our results, or do
not have adequate resolution.

\section{Discussion and Conclusions}
\label{conclusions}

\begin{figure}[t!]\begin{center}
\includegraphics[width=\columnwidth]{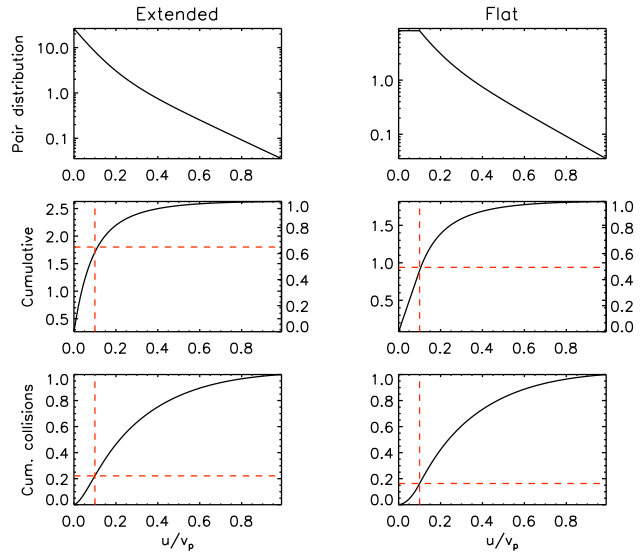}
\end{center}\caption{Ad hoc fits for $u<u_c$.  Left column: extending \Eq{fit!} to $u=0$, Right column:
extending the constant value at $u=u_c$ to $u=0$.
Top panels: pair density, i.e., \Eq{fit!}.
Middle panels: solid: cumulative totals of the top panels; dashed: $u_c$ its intercepts by the data.  The right axis is
self-normalized. 
Bottom panels: solid: self-normalized cumulative totals of the collision rate i.e., the first velocity moment of \Eq{fit!};
dashed: $u_c$ its intercepts.
\label{extensions} }
\end{figure}

We arrive at a picture of turbulence-induced particle concentration and collisions which creates spatially small (with
respect to turbulent scales),
highly correlated ``cold'' clusters of particles.  While these clusters are themselves non-collisional, they periodically
pass through each other, resulting in bursts of collisions.  These clusters are channeled through the high-strain
boundaries between turbulent eddies, and even separate clusters are strongly correlated.  This correlation
causes the turbulent collision speeds we find to be significantly slower than those predicted by analytical
approaches, that cannot treat the long time-scale correlations.
This picture of particle-particle collisions
occurring when clusters of particles pass through each other
 is supported by the long stabilization time
seen in \Fig{cluster}, where it takes many turbulent turnover times to reach a statistical steady state, implying non-trivial dust spatial structures.

Our results apply most strongly for particles deeply embedded in a turbulent cascade.
Our approach does not have the ability to handle the details of the forcing scale or the dissipation scale of the turbulence,
both of which could have different behavior such as the bottleneck effect at the dissipation scale \citep{D03}.
Nevertheless, the qualitative similarity between our low inertial range runs
(Basic, Q-A, P-A, G-A) with the large inertial range runs (B-LI2, Q-LI, P-LI, G-LI) suggests that this will only pose a difficulty
if the statistics of the turbulence are indeed quite different at those two scales.  A second difficulty is that our synthetic turbulence
model cannot handle the dust's back-reaction on the gas, which can be significant if there is strong clustering.  An implicit assumption
of our work therefore is that the dust spatial density is less than that of the gas.

We find that \Eq{fitform} is a workable fit for the number of turbulence-induced particles pairs as a function of collisional velocity.
This form has some interesting implications.  A simple one is that it is broad: unlike the case of, for example, a Maxwellian
distribution, enough of collisional velocity space is populated that many collisional outcomes will occur
in relevant quantities.  Less obviously, we note that in the ``high'' velocity parameter regime,
we can fit the collisional probability distribution through the autocorrelation of exponential decays:
\EQ
C_b ue^{-\frac{u}{u_b}}=\int_0^{u}\left[\sqrt{C_b} e^{-\frac{u'}{u_b}}\right]\left[\sqrt{C_b}e^{-\frac{u-u'}{u_b}}\right] du'.
\EN
This implies that (faster) collisions are generated by particles each contributing a velocity with respect to the collisional
center (which is a feature of the fluid flow rather than the center of mass of the particle pair)
with an exponentially falling probability.  Note that this is not the difference between the particle velocity and that
of the gas, which is, naturally, correlated between the two particles.  Why the individual particle probability distribution
is an exponential decay, and why the autocorrelation occurs in collision-space ($P(u) \propto ue^{-u/u_0}$) rather
than in the particle pair space ($P(u) \propto e^{-u/u_0}$) is unclear.

Our formula is limited in its applicability to $u>u_c$, and as $R$ tends to $0$, neglects almost all the particle pairs
since they are part of the cold population (because $\mu>0$, see the height of the peak in the middle panel of \Fig{paircounts},
which increases as $R$ decreases).
This does not appear to pose unsurpassable difficulties: in \Fig{extensions} we show two alternate ad-hoc extensions
for \Eq{fit!} into the low $u$-regime.  From the bottom panels we can see that it is unlikely that
errors in the collision rate based on the extension chosen will be more than $25\%$ of the total: low relative velocity dust grains,
even if numerous, contribute few collisions due to their slow speeds.  Errors in energy,
the square-root of the 3rd velocity moment, will be significantly less yet, as long as there is not a hidden, highly clustered but
still collisional population that we have not been able to detect so far.

Another intriguing result is the low scale of $u_b\simeq 0.2v_p$.  This collisional velocity scale
is significantly below the predicted rms collisional velocity $u_{\text{col}}=2^{1/2} v_p$ of 
\cite{OC07}.  This might be explained by the vertical axes of the middle plots of \Fig{extensions}: the limiting value
on the left axes gives the effective concentration i.e., the number of pairs in the warm and hot populations, divided by $\overline{N}(R)$.
While the difficulties fitting the cold population make the result uncertain, it appears that 
clustering occurs in the collisional population: particles see more warm and hot collisional partners than would
be expected if all dust grains were homogeneously distributed in space.
This clustering is different from that of the cold population in that
the clustering is independent of $R$ for small separations (i.e., the effective $\mu=0$; an equivalent plot to the
bottom panel of \Fig{cluster} would be flat).
For any clustering to occur, the colliding particles in the warm and hot populations must
still be significantly correlated with one another, an aspect which the work of \citet{OC07} could not fully capture.

In conclusion, we see turbulence-induced dust collisions at significantly (a factor of 3-5) slower velocities
than predicted by existing analytical theory.
If we consider an approximate $\alpha$-disk Minimum Mass Solar Nebula with a sound speed of $c_s=10^5\,$cm/s (temperature
$T \simeq 300\,$K) at $1\,$AU, $\alpha=10^{-3}$ and centimeter-scale grains with a stopping time $\tau_p \Omega_K=0.01$ and
mass $m\sim 1\,$ g, the turbulent
velocity at the largest scale is $v_t = 3 \times 10^3\,$cm/s \citep{SS73,Hayashi81}.  Under these conditions, the predicted collisional velocity of \cite{OC07} is
$u=5 \times 10^2\,$ cm/s, well into the destructive collision regime of \cite{G10}, Figure 11.  A five-fold reduction in the collision
speed would however lead to bouncing.  While this result implies frequent destructive collisions, we also
predict large numbers of lower velocity collisions, which could lead to sticking.
This would ease the difficulties in planetesimal formation associated with 
bouncing and fragmentation \citep{Z10}.  The velocities we expect will still lead to significant fragmentation,
but the inclusion of a slower collisional velocity probability distribution allows for the consideration of ``lucky'' particles
that are in unusually low velocity collisions to grow large enough that they can survive. 
 While the slower collisional velocities reduces collision rates, we also see
a limited enhancement of the effective dust number density through clustering, which mitigates this collision rate reduction. 

Finally, the hot population collisional tail falls off only exponentially with velocity, rather
than something closer to a Maxwellian distribution.  This emphasizes the importance of using a collisional
velocity probability distribution instead of a single characteristic collisional velocity.
For example, high velocity outlying events are expected to occur at non-negligible rates, and could contribute to a fragmentation cascade
and dust reprocessing.  This would occur even if the reduction in collisional velocities results in few enough destructive collisions that the fragmentation barrier to dust growth is lifted.

Our results are well suited to inclusion in a model of collisional dust grain agglomeration in protoplanetary disks such
as that of \cite{Z10}.  The formula
given by \Eq{fitform} is simple enough for inclusion, while allowing a velocity probability distribution.
This enables the full use of experimental results about the critical velocities at which colliding particles stick, bounce or fragment.

\appendix

\section{Stokes numbers}
\label{compstokes}

Studies of turbulent particle transport generally use the Stokes number $St \equiv \tau_p/\tau_{\eta}$ to non-dimensionalize
the stopping time, where $\tau_{\eta}$ is the turbulent time scale associated with the viscous dissipation scale.
Astrophysical studies of protoplanetary disks however often use $St \equiv \tau_p \Omega_K$, where 
$\Omega_K$ is the Keplerian rotation rate, since it is better constrained than $\tau_{\eta}$ and is believed to be a good
estimate for the time scale associated with the largest scale turbulence.  As such, the particle stopping time is often
scaled to two completely different time scales, the largest and the smallest associated with the turbulence.
Neither formulation is however clearly appropriate for studies of particle-particle
relative motion and clustering.

Stating that the relevant
quantity is $St= \tau_p/\tau_{\eta}$ implies that either the details of the dissipation process
proper, or the lack of fluid energy at scales $k>k_{\eta}$ plays a crucial role in the particle response
to the turbulence.  For a particle with $St \lesssim 1$, such effects are expected, as particles
couple most strongly with eddies with turnover times $t\sim \tau_p$ and the dissipative cutoff
means that many of those eddies are missing.
However, it is unclear why a dependence on $\tau_p/\tau_{\eta}$ would exist
for particles with $St \gg 1$.  Instead,
for particles with $\tau_{ls} \gg \tau_p \gg \tau_{\eta}$, we expect the behavior to be scale free: the largest
 scale of the turbulence ($\tau_{ls}$) is too large to affect particle-particle relative motion while
the dissipation scale is too small, so the turbulence does not set time or length scales
for the particles.  In this regime, instead of measuring the particle stopping time as a function
of some turbulent time, one should instead measure the turbulence by the particle stopping time.
This approach is implicitly followed by \cite{V80}, where the division of eddies into Classes I
and II is done by measuring the turbulent turnover time relative to $\tau_p$.  They do
quote results as a function of $St \equiv \tau_p \Omega_K$, but that is only possible because
of the assumption of a Kolmogorov cascade, which allows them to note that eddies with turnover times
 $t=\tau_p$ have velocities $v_p=\sqrt{St} v_{ls}$.

This poses difficulties in numerical simulations of particle relative motion in turbulence because
the accessible ratio of $\tau_{ls}/\tau_{\eta}$ is modest at best.  Accordingly, any apparent dependence
on $St\equiv \tau_p/\tau_{\eta}$ is difficult to distinguish from a dependence on $St\equiv \tau_p/
\tau_{ls}$ and even then would imply little for the case of a particle deeply embedded in a
large cascade.  For example, when considering the work of \cite{P11},
it should be noted that their ability to track the clustering of particles
deeply embedded in a turbulent cascade ($\tau_p \gg \tau_{ss}$ but also $\tau_p \ll \tau_{ls}$) is
limited by their modest inertial range.  Alternatively, Figure 3 of \cite{Carballido10} plots
the difference between theoretical results for a large inertial range (solid) and the theoretical
results applied to the numerically obtained inertial range (dotted).  Any extension of
their results of decreased clustering for large $St$ is suspect for a protoplanetary disk with
a large inertial range.  It does, however, certainly imply that clustering will decrease once
$\tau_p /\tau_{ls} \simeq \tau_p \Omega_K$ becomes large.


\begin{table}[b!]
\caption{Variables, Parameters an Diagnostics \label{var}}
\centerline{\begin{tabular}{lll}
\hline \hline
Name & Notes & Section\\
\hline
\hline
$\mmm$ & Shell numbering for largest scale shell &  \multirow{3}{0.2\linewidth}{\Sec{pureKol}}  \\
$\mMM$ & Shell numbering for smallest scale shell & \\
$\mmp$ & Shell numbering for shells with $k=k_p$ & \\
\hline
$ls$ & Subscript for large-scale, equivalent to $\mMM$  & \multirow{2}{0.2\linewidth}{\Sec{pureKol}} \\
$ss$ & Subscript for small-scale, equivalent to $\mmm$ & \\
$p$ & Subscript for particle, equivalent to $\mmp$ & \Sec{Sec:Particles} \\
\hline
$\VV(\xx,t)$ & Gas velocity & \Eq{vgmodel}\\ 
$\tau_{ls}$ &Turbulent turnover time at the largest turbulent scale & \multirow{2}{0.2\linewidth}{\Sec{pureKol}} \\
$\tau_{ss}$ & Turbulent turnover time at the smallest turbulent scales &   \\
\hline
$\kk_{mn}$ & \multirow{5}{0.4\linewidth}{Synthetic turbulence parameters} & 
                         \multirow{5}{0.2\linewidth}{\Sec{pureKol}} \\
$\hat{\vv}_{mn}$ & & \\
$v_{mn}$ & & \\
$a_{mn}$ & & \\
$\phi_{mn}$ & & \\
\hline
$\uu_p$ & Particle velocity & \multirow{5}{0.2\linewidth}{\Sec{Sec:Particles}} \\
$\tau_p$ & Particle stopping time &   \\
$k_p $ & Wavenumber of turbulence with turnover time $\tau=\tau_p$ & \\
$v_p$ & Gas velocity at $k=k_p$ & \\
$\Steff$ & Effective Stokes number & \\
%
\hline
$u_a$ & \multirow{5}{0.4\linewidth}{Coefficients for the particle-pair number and velocity probability distribution fit} &
                \multirow{5}{0.2\linewidth}{\Eqs{fitform}{uc}} \\
$u_b$ & & \\
$u_c$ & & \\
$C_a$ && \\
$C_b$ & & \\
\hline
$N(R,u,t)$ & Particle pair count within separation R, at velocity $u$ in snap-shot $t$  & \multirow{7}{0.2\linewidth}{\Sec{IA-clustering}}  \\
$N(R,u)$ & as $N(R,u,t)$, averaged over snap-shots & \\
$N(R,t)$ & as $N(R,u,t)$, integrated over velocity & \\
$N(R)$ & as $N(R,u,t)$, integrated over velocity and averaged over snapshots & \\
$C(R)$ & Particle concentration &  \\
$\mu$ & Power-law exponent for C(R) & \\
$R_1$ & $y=1$ intercept for power-law fit of C(R)  & \\
\hline
$N'(R,u)$ & Pair density in velocity space: $N(R,u)/\triangle u$ & \Eq{Nprime} \\
$\omega$ & Cold population velocity measure & \Eq{omega} \\
$R_c$ & Contamination radius & \Eq{radcont} \\
\hline \hline
\end{tabular}}
\end{table}

\acknowledgements

This work was supported by  a fellowship from the Alexander von Humboldt Foundation. Further support was provided by 
a Kalbfleisch Fellowship
from the American Museum of Natural History and the NSF through CDI grant AST08-35734 and AAG grant AST10-09802.
We thank Dhrubaditya Mitra for his kind advice and providing the GOY routines.
The computations have been carried out at the
National Supercomputer Centre in Link\"oping and the Center for
Parallel Computers at the Royal Institute of Technology in Sweden.


\end{document}